\documentclass[fleqn,usenatbib]{mnras}

\usepackage{amsmath}
\usepackage{amsfonts}
\usepackage{graphicx}
\usepackage{aas_macros}
\usepackage{color}
\usepackage{listings}
\usepackage{hyperref}

\begin{document}

\title{Bayesian cross validation for gravitational-wave searches in pulsar-timing array data}

\author[Haochen Wang et.\ al]{Haochen Wang,$^{1}$ Stephen R. Taylor,$^{2}$ and Michele Vallisneri$^{3}$\\ \\
$^{1}$Department of Physics and Astronomy, University of Southern California, Los Angeles, California 90089, USA\\
$^{2}$TAPIR, MC 350-17, California Institute of Technology, Pasadena, California 91125, USA\\
$^{3}$Jet Propulsion Laboratory, California Institute of Technology, Pasadena, California 91109, USA
}

\maketitle

\begin{abstract}
Gravitational-wave data analysis demands sophisticated statistical noise models in a bid to extract highly obscured signals from data.
In Bayesian model comparison, we choose among a landscape of models by comparing their marginal likelihoods. However, this computation is numerically fraught and can be sensitive to arbitrary choices in the specification of parameter priors.
In Bayesian cross validation, we characterize the fit and predictive power of a model by computing the Bayesian posterior of its parameters in a training dataset, and then use that posterior to compute the averaged likelihood of a different testing dataset. The resulting cross-validation scores are straightforward to compute; they are insensitive to prior tuning; and they penalize unnecessarily complex models that overfit the training data at the expense of predictive performance.
In this article, we discuss cross validation in the context of pulsar-timing-array data analysis, and we exemplify its application to simulated pulsar data (where it successfully selects the correct spectral index of a stochastic gravitational-wave background), and to a pulsar dataset from the NANOGrav 11-year release (where it convincingly favors a model that represents a transient feature in the interstellar medium).
We argue that cross validation offers a promising alternative to Bayesian model comparison, and we discuss its use for gravitational-wave detection, by selecting or refuting models that include a gravitational-wave component.
\vspace{12pt}
\end{abstract}

\section{Introduction}

Searches for gravitational waves (GWs) in pulsar-timing-array (PTA) data \citep{lommen15,bs15} seek to identify weak GW signals among a plethora of other effects, including deterministic delays due to the relative motion of pulsar and observatory and to pulsar binary dynamics, stochastic delays due to the interplanetary and interstellar media, as well as intrinsic irregularities in the pulsar's period emission \citep{Cordes_2013,Stinebring_2013}. These searches are commonly formulated as Bayesian-inference problems \citep{gregory10}, whereby we derive the joint posterior probability density of the GW parameters and of the noise parameters of all analyzed pulsars. Choosing appropriate probabilistic models for pulsar noise is therefore crucial to reliable PTA searches \citep{2013PhRvD..87d4035T,2016MNRAS.458.2161L,2010arXiv1010.3785C}: unmodeled noise components may be interpreted as GWs, while overgenerous noise assumptions may reduce GW sensitivity.
In current practice, pulsar noise models are informed by the physics of millisecond pulsars and of the interplanetary/interstellar medium, but they are largely driven by inference from PTA datasets, since these often represent the best observations to date for PTA pulsars.

\section{Model comparison}

Within the data-analysis practice of the NANOGrav collaboration \citep{nanograv}, \emph{Bayesian model comparison} \citep{gregory10} is used to select the noise model most appropriate to each pulsar \citep{nano_noise}. The goal is not only to improve the physical characterization of the processes affecting pulse times of arrival (TOAs), but also to isolate these processes from a putative GW signal with greater confidence. In this framework, we evaluate the fully marginalized likelihood (a.k.a.\ evidence) for each model $M$:
\begin{equation}
\label{eq:evidence}
p(y|M) = \int p(y|\theta_M) p(\theta_M) \, \mathrm{d} \theta_M,
\end{equation}
where $y$ denotes the observed data, $\theta_M$ the parameters of model $M$, $p(y|\theta_M)$ the likelihood (the probability of $y$ given $\theta_M$), and $p(\theta_M)$ the prior probability density assigned to the parameters.
We then compare models by evaluating Bayes ratios\footnote{Bayesian model comparison calls for the computation of \emph{odds ratios}, which account for the prior relative probability of entire models. Since it is very difficult to attribute such priors on physical grounds, we generally work directly with Bayes ratios.} $B_{21} = p(y|M_2) / p(y|M_1)$, either directly through Eq.\ \eqref{eq:evidence} (often requiring significant numerical sophistication, see \citealt{2008ConPh..49...71T}) or by the Monte Carlo exploration of uber-likelihoods that specialize to individual models depending on the value of an index parameter (in which case the Bayes ratio is given by the ratio of the ``time'' spent in each model, see \citealt{godsill2001relationship,sisson2005transdimensional}).
A large Bayes ratio $B_{21}$ implies that the data favors model $M_2$ over $M_1$. However, it is difficult to give Bayes ratios a principled \emph{quantitative} interpretation. The exception are cases where alternative models represent exclusive physical outcomes; Bayes ratios can then be calibrated in terms of statistical decision theory, by relating their sampling distribution to false-alarm and false-dismissal probabilities (see, e.g., \citealt{v12}).

In choosing between alternative models for a dataset, we need to be wary of \emph{overfitting}: that is, while it is always possible to improve model fit by adding parameters, the enhanced model may end up conforming to contingent noise features instead of highlighting the physical properties of interest. Correspondingly, the model loses predictive power for yet-to-observed data. Bayesian model comparison incorporates a defense against overfitting, in that the evidence integral penalizes fine tunings that restrict parameters to small regions within their prior ranges.
Unfortunately, this defense creates a different, significant weakness---Bayes ratios are then sensitive to parameter-prior assignments that may be largely arbitrary, and are not testable from data \citep{g13}. Consider for instance the case of two nested models $M_1 \subset M_2$, where $M_2$ is obtained by adding parameter $\theta^*$ to $M_1$; let $\theta^*$ have uniform prior in $[-a,a]$, with $a$ arbitrarily assigned without cogent physical grounds; finally, let the data constrain $\theta^*$ to a small range close to $0$. It is then easy to see that $B_{21} \propto 1/a$.

\section{Cross validation}

An alternative framework for model comparison is offered by measures of \emph{predictive performance}, which quantify how well a model that has been fit to dataset $y$ can predict yet-to-be-observed data $\tilde{y}$ \citep{g13}. These measures penalize overfitting by construction, because a model that conforms to contingent features in $y$ will usually do worse in fitting $\tilde{y}$. In a Bayesian setting, a commonly adopted measure of predictive performance is the \emph{log predictive density} for the new data $\tilde{y}$, as induced by the posterior $p(\theta_M|y)$:
\begin{equation}
\log p(\tilde{y}|y;M) = \log \int p(\tilde{y}|\theta_M) p(\theta_M|y) \, \mathrm{d}\theta_M.
\end{equation}
Ideally, we would average $\log p(\tilde{y}|y;M)$ over the true distribution of future data $\tilde{y}$; doing so is however seldom possible. In practice, we can: a) estimate \emph{within-sample} predictive accuracy using the data $y$ that we already have, by applying corrections for the overfitting bias, as in the various ``information criteria'' \citep{g13}; b) evaluate \emph{out-of-sample} predictive accuracy on one or more \emph{holdout} datasets that were not used to infer parameter posteriors. The latter approach is known as \emph{cross-validation}, and we will pursue it for PTA data in the rest of this article. 

Specifically, we adopt \emph{$k$-fold} cross validation as follows:
\begin{enumerate}
\item We divide the dataset $y$ randomly in $k$ exclusive subsets $y^{(k)}$ (the \emph{testing} datasets); 
\item For each $k$, we derive the posterior $p(\theta_M | y^{(-k)})$, where $y^{(-k)} = \cup_{j \neq k} y^{(j)}$ is the \emph{training} dataset corresponding by omitting $y^{(k)}$. We represent posteriors as sequences $\{\theta^{(-k)}_{M,i}\}$ of $N$ quasi-independent samples, obtained by Monte Carlo methods;
\item For each $k$, we evaluate the log predictive density $\log p(y^{(k)}|y^{(-k)})$, given by
\begin{multline}
\label{eq:lpd}
\log \int p(y^{(k)}|\theta_M) p(\theta_M|y^{(-k)}) \, \mathrm{d}\theta \\ \simeq \log \frac{1}{N} \sum_{i=1}^N p(y^{(k)}|\theta^{(-k)}_{M,i});
\end{multline}
\item We repeat this procedure for every model under consideration, and then compare the respective log predictive densities, averaged over the $k$ repetitions. The variance of the densities is a measure of their statistical uncertainty.
\end{enumerate}

\begin{figure} 
\includegraphics[width=\columnwidth]{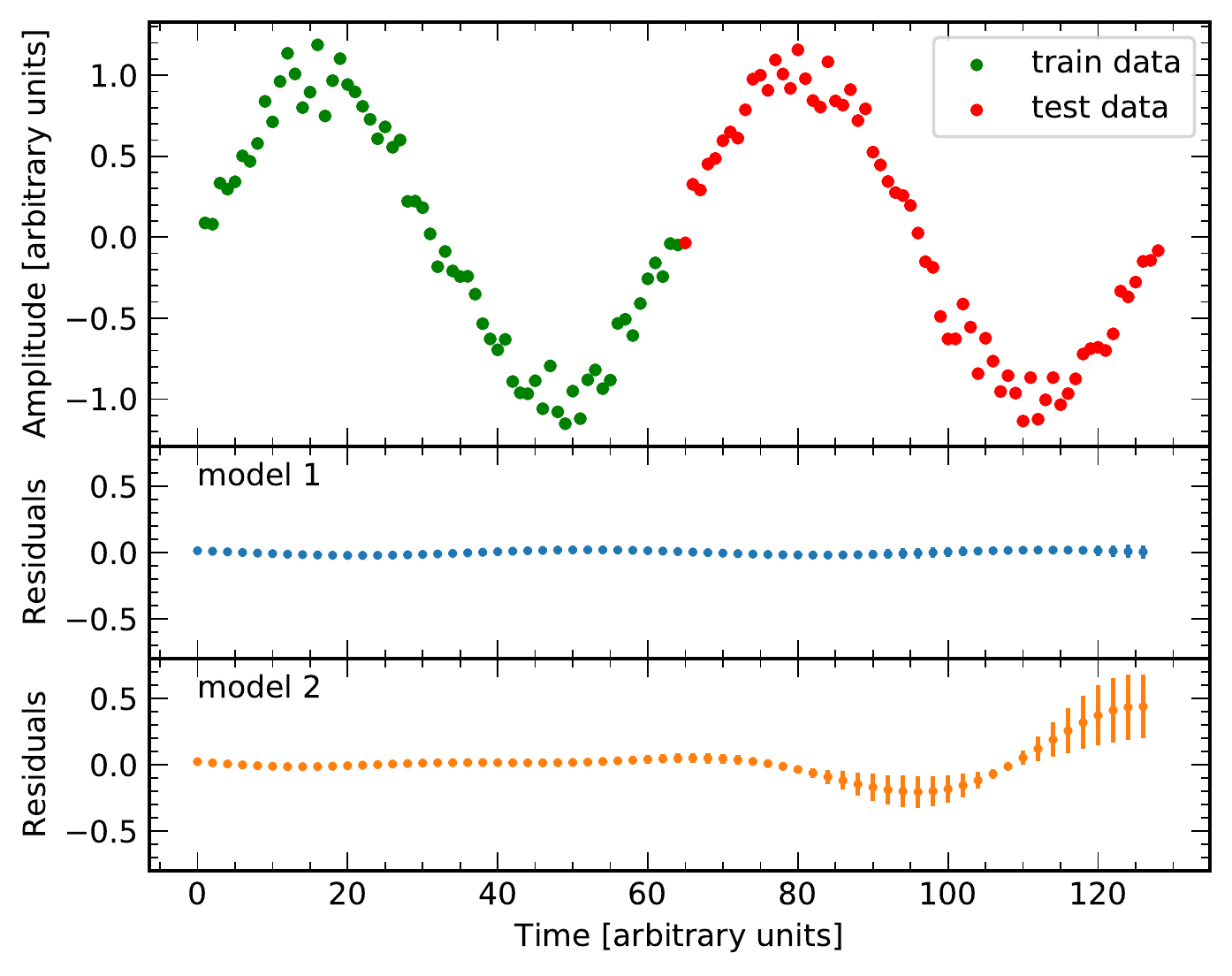}
\caption{Cross-validation analysis of a toy model.
We generate the data $d$ (upper panel) from $f_1(A,\omega,\varphi)=A \sin(\omega t+\varphi)$ (model 1), with the addition of white noise; we then analyze $d$ with $f_1$ as well as $f_2(A,\omega,\varphi,b)=A \sin((\omega + b t) t+\varphi)$ (model 2), where parameter $b$ describes an additional frequency drift.
We train both models on the first half of the data (the training set $d^\mathrm{train}$, green in the plot), deriving the Bayesian posterior distributions $p(A,\omega,\varphi|d^\mathrm{train},f_1)$ and $p(A,\omega,\varphi,b|d^\mathrm{train},f_2)$. Using those posteriors with Eq.\ \eqref{eq:lpd} over the second half of the data (the validation set $d^\mathrm{test}$, red in the plot), we obtain log predictive densities $-27.7$ and $-31.3$ for models 1 and 2 respectively. The lower density for model 2 indicates that it overfits the training data, resulting in a poor fit to the validation set. The conclusion is borne out by inspection of model residuals (lower panels), represented in the plot as their averages and standard deviations over 300 posterior draws. Model-1 residuals are small and homogeneous across both training and validation sets; model-2 residuals have much larger bias and variance over the validation set, explaining the lower predictive density.}
\label{fig1}
\end{figure}

In \autoref{fig1}, we exemplify this process with a simple toy problem: a sinusoidal signal parametrized by amplitude, frequency, and phase, as modeled by that very model and by an expanded model that includes a linear frequency drift. The more complicated model results in significantly lower log predictive density, demonstrating that cross validation can recognize and reject overfitting.

\section{Cross validation for single-pulsar noise modeling}

In current practice (see \citealt{vv14} for a recent review), probabilistic noise models for pulsars are built as the sum of a number of Gaussian processes (GPs; \citealt{2006gpml.book.....R}) representing all sources of correlated noise: errors in the parameters of the deterministic timing model, pulsar-rotation irregularities, dispersion-measure variations along the pulse propagation path, jitter-like noise in multifrequency observations, and more. The TOAs are also subject to ``white'' radiometer measurement noise, conceptualized as independent and heteroskedastic normal variates. Both the GPs and measurement noise are governed by a set of hyperparameters (e.g., the amplitude and spectral slope of delays due to rotation irregularities) that are estimated from PTA datasets.

In keeping with basis--kernel duality for GPs, the model likelihood can be written in two complementary ways. In \emph{hierarchical form}, the likelihood is given by
\begin{multline}
\label{eq:hierarchical}
    p(y|\eta_N,\eta_\mathrm{GP},c_\mathrm{GP}) = p(y|\eta_N,c_\mathrm{GP}) \times p(c_\mathrm{GP}|\eta_\mathrm{GP}) \\ = 
    \frac{\mathrm{e}^{-(y - F_\mathrm{GP} c_\mathrm{GP})^T N^{-1} (y - F_\mathrm{GP} c_\mathrm{GP})/2}}{\sqrt{(2\pi)^n |N|}}
    \times
    \frac{\mathrm{e}^{-c_\mathrm{GP}^T \Phi_\mathrm{GP}^{-1} c_\mathrm{GP}/2}}{\sqrt{(2\pi)^m |\Phi|}},
\end{multline}
where $y$ is the vector of $n$ \emph{timing residuals} obtained by subtracting the best-fit timing model from the observed pulse times of arrival; the $n \times m$ matrix $F_\mathrm{GP}$ collects the $m$ GP basis vectors, and the $c_\mathrm{GP}$ are the corresponding weights (or coefficients); $N$ (a function of the hyperparameters $\eta_N$) is a diagonal matrix expressing measurement-noise variance; and $\Phi_\mathrm{GP}$ (a function of the hyperparameters $\eta_\mathrm{GP}$) represents the normal priors for the GP weights. In \emph{marginalized form}, we eliminate the dependence on the GP weights by integrating over them \citep{vv14}:
\begin{multline}
\label{eq:marginalized}
    p(y|\eta_N,\eta_\mathrm{GP}) = \int p(y|\eta_N,c_\mathrm{GP}, \eta_\mathrm{GP}) \, \mathrm{d}c_\mathrm{GP} \\ =
    \frac{\mathrm{e}^{-y^T (N + F_\mathrm{GP} \Phi_\mathrm{GP} F_\mathrm{GP}^T)^{-1} y/2}}{\sqrt{(2\pi)^n |N + F_\mathrm{GP} \Phi_\mathrm{GP} F_\mathrm{GP}^T|}}.
\end{multline}
The marginalized form is usually employed for the Monte Carlo exploration of hyperparameter posteriors. The GP weights can still be characterized by way of their conditional posterior given the data and the hyperparameters,
which follows the jointly normal distribution $p(c_\mathrm{GP}|y;\eta_N,\eta_\mathrm{GP}) = \mathcal{N}(\bar{c},\Sigma)$ with mean
\begin{equation}
\label{eq:conditionalmean}
\bar{c}(y;\eta_N,\eta_\mathrm{GP}) = \Sigma F^T_\mathrm{GP} N^{-1} y
\end{equation}
and covariance
\begin{equation}
\label{eq:conditionalcov}
\Sigma(\eta_N,\eta_\mathrm{GP}) = (\Phi_\mathrm{GP}^{-1} + F_\mathrm{GP}^T N^{-1} F_\mathrm{GP})^{-1}.
\end{equation}

Armed with Eqs.\ \eqref{eq:hierarchical}--\eqref{eq:conditionalcov}, we perform steps 1--3 of $k$-fold cross-validation for a single-pulsar dataset as follows:
\begin{enumerate}
    \item We partition the timing residuals $y$ into exclusive testing datasets $y^{(k)}$, making sure that each training dataset $y^{(-k)}$ depends on all weights and hyperparameters. For instance, for dispersion-measure variations described as piecewise-constant ``DMX'' functions, each DMX epoch must be populated by at least one residual in every $y^{(k)}$; likewise, for ``EFAC'' measurement noise that is rescaled differently in each radio backend, each backend must be represented in every $y^{(k)}$;
    \item We sample the marginalized hyperparameter posterior $p(\eta_N,\eta_\mathrm{GP}|y^{(-k)})$ [proportional to Eq.\ \eqref{eq:marginalized} times prior $p(\eta_N,\eta_\mathrm{GP})$], using the PTA data-analysis package \texttt{Enterprise}\footnote{\href{https://github.com/nanograv/enterprise}{https://github.com/nanograv/enterprise}} and the Markov Chain Monte Carlo sampler \texttt{PTMCMCSampler};\footnote{\href{https://github.com/jellis18/PTMCMCSampler}{https://github.com/jellis18/PTMCMCSampler}}
    \item For each of the $N$ quasi-independent $(\eta^{(-k)}_{N,i},\eta^{(-k)}_{\mathrm{GP},i})$ obtained at step 2, we draw $P$ weight vectors $\{c^{(-k)}_{\mathrm{GP},ij}\}$ from their conditional distribution [Eqs.\ \eqref{eq:conditionalmean}--\eqref{eq:conditionalcov}], then we evaluate the log predictive density by averaging the hierarchical likelihood \eqref{eq:hierarchical} over the $N \times P$ triples $(\eta^{(-k)}_{N,i},\eta^{(-k)}_{\mathrm{GP},i},c^{(-k)}_{\mathrm{GP},ij})$.
\end{enumerate}

The ``representation'' condition imposed in step 1 is necessary so that the parameters for which we derive posteriors at step 2 fully specify the model's prediction for each testing dataset; this prediction is used in step 3 to evaluate the predictive density. Another important technical subtlety is that the GP weights must conserve the same identity across all $y^{(-k)}$ (for example, Fourier coefficients for the same set of frequencies in the case of correlated spin noise); by contrast, the GP basis vectors would change in value, because they refer to different time-of-arrival measurements (continuing our example, the basis elements would be sines and cosines of the same frequencies for all subdatasets, but would be evaluated at different times).

We note also that in step 2 we could have used the hierarchical likelihood to sample the full parameter set $(\eta_N,\eta_\mathrm{GP},c_\mathrm{GP})$, avoiding the conditional $c_\mathrm{GP}$ draws at step 3. However, the hierarchical likelihood is considerably harder to explore stochastically. Also, in step 3 the sum over the $c^{(-k)}_{\mathrm{GP},ij}$ for each $i$ could be replaced by analytical integration in terms of $\bar{c}$ and $\Sigma$, at the cost of some algebraic complication (see below for a related development).

\section{Results}

\begin{figure}
\includegraphics[width=\columnwidth]{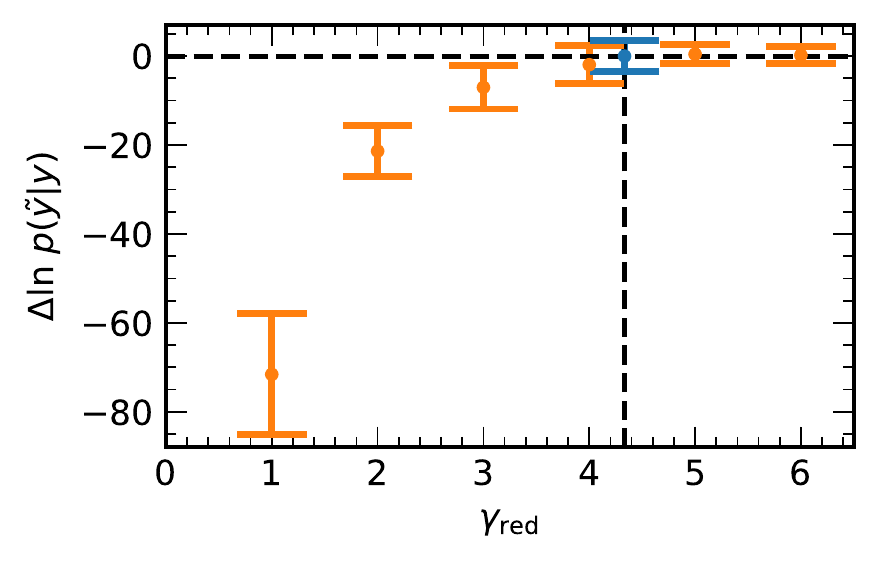}
\caption{Cross-validation analysis of power-law models for pulsar red noise, as demonstrated in the J1713+0747 dataset from the $1^\mathrm{st}$ IPTA Mock Data Challenge.
The dataset includes GW-like correlated noise with spectral slope  $\gamma_\mathrm{red} = 13/3$. We perform two-fold cross validation using power-law models with integer $\gamma_\mathrm{red} \in [1,6]$, as well as the correct $\gamma_\mathrm{red} = 13/3$.
For each $\gamma_\mathrm{red}$, we plot the average and standard deviation of the log predictive density over five random shuffles of the data into training and validation subsets. We adopt the standard deviation as a proxy for the uncertainty of the predictive density. All values are shown relative to the $\gamma_\mathrm{red} = 13/3$ result (plotted in blue). Lower values of the spectral slope are clearly disfavored, while the dataset cannot discriminate among slopes $\gamma_\mathrm{red} > 4$, for which the characteristic correlation timescale of red noise exceeds the span of the measurements.}
\label{fig2}
\end{figure}

To demonstrate how cross validation can be applied to PTA noise-model selection, we first consider the simulated TOA residuals for pulsar J1713+0747 from the $1^\mathrm{st}$ International Pulsar Timing Array (IPTA) Data Challenge \citep{2018arXiv181010527H}.\footnote{\href{http://ipta4gw.org/data-challenge}{http://ipta4gw.org/data-challenge}} These residuals (130 values at 14-day cadence) are generated from a simplified deterministic timing model and a simple noise model: the timing model parameters describe the intrinsic-spin, astrometry, and binary-orbit properties of the pulsar; the noise model includes white measurement noise (described by ``EFAC'' and ``EQUAD'' parameters) and red correlated spin noise, specified by a power-law spectrum \citep{arz18b}
\begin{equation}
P(f) = A_\mathrm{red}^2\left(\frac{f}{f_\mathrm{yr}}\right)^{-\gamma_\mathrm{red}},
\end{equation}
where \(A_\mathrm{red}\) is the amplitude of the red-noise process in units of \(\mu s \times \mathrm{yr}^{1/2}\), \(\gamma_\mathrm{red}\) is its spectral index (set to 13/3 to generate the data), and \(f_\mathrm{yr} = 1\,\mathrm{yr}^{-1}\). We compare power-law red-noise models with different \(\gamma_\mathrm{red}\) by evaluating their respective cross-validation predictive densities, shown in \autoref{fig2} for \(\gamma_\mathrm{red}\) ranging from 1 to 6. It is clear that a value larger than 4 (and consistent with 13/3) is preferred.

\begin{figure}
\centering
\includegraphics[width=0.5\textwidth]{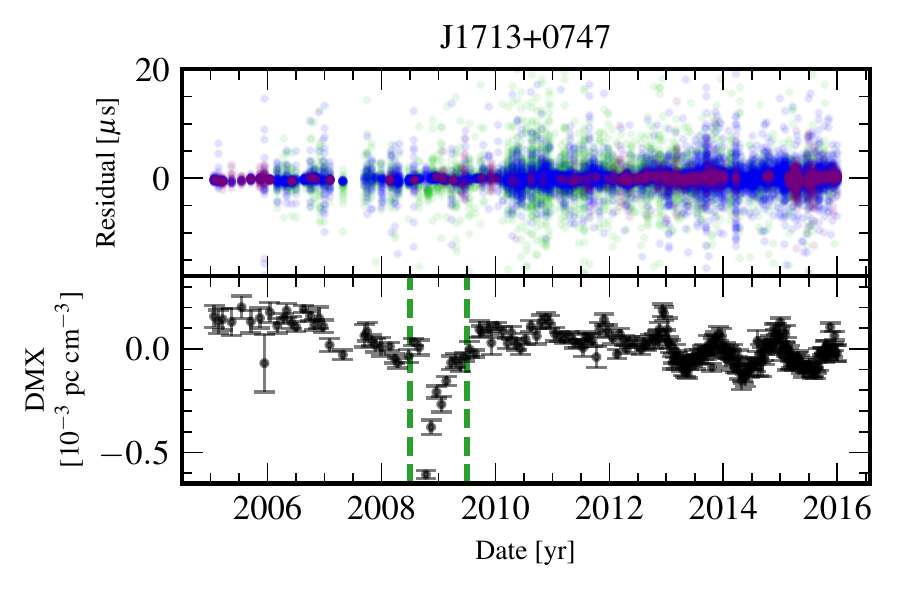}
\caption{Residuals and dispersion-measure variation (DMX) for pulsar J$1713$+$0747$ in the NANOGrav 11-yr data release \citep{arz18a}. DMX shows a dramatic dip around year 2009. The dashed green vertical lines on the lower panel indicate a one-year window centered around 2009. Colors in the upper panel indicate observations taken at different radio frequencies: Blue: 1.4 GHz; Purple: 2.1 GHz; Green: 820 MHz.}
\label{fig3}
\end{figure}

Moving on to real PTA datasets, we perform two-fold cross validation on the TOA residuals of pulsar J1713+0747 from the NANOGrav 11-yr data release \citep{arz18a}. As apparent in \autoref{fig3} (from \citealt{arz18a}), around 2009 the residuals underwent a dispersion-measure (DM) dip (i.e., an apparent decrease in the electron density experienced by radio pulses traveling to Earth, resulting in reduced ``fanning'' across frequencies).
We compare two noise models: the first (\emph{GP+dip}) represents DM with a Gaussian process, but it includes also a transient feature with exponential decay to represent the dip; the second (\emph{GP-only}) represents DM with the Gaussian process alone. In keeping with the representation condition introduced above, when we select the training dataset we need to make sure that it includes a sufficient number of residuals around the dip. To achieve this, we build the training dataset by randomly selecting half of the residuals within a one-year window centered around 2009 (see \autoref{fig3}), as well as half of the residuals outside the window.

\begin{figure}
\centering
\includegraphics[scale=0.85]{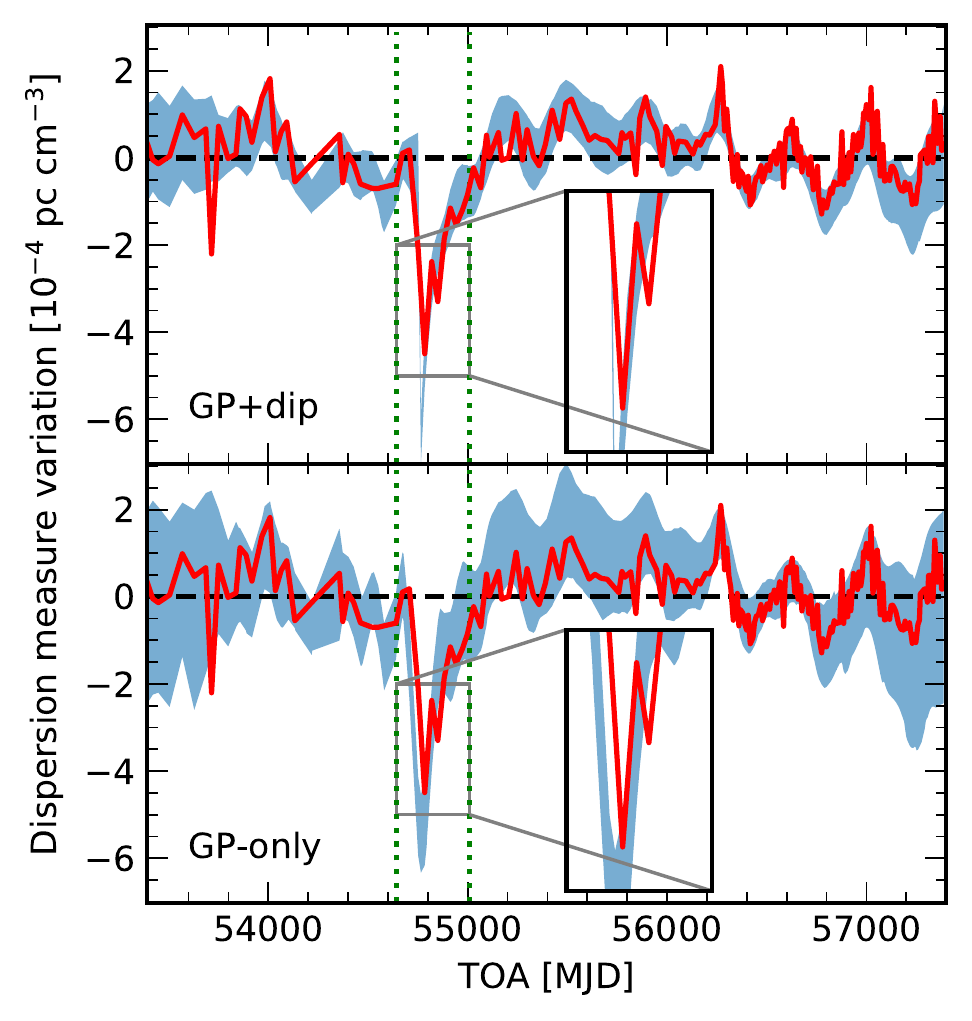}
\caption{Reconstructed dispersion measure according to the GP+dip model (upper panel) and the GP-only model (lower panel), shown as the 5\% to 95\% posterior interquantile range (gray bands) in a single cross-validation run. ``True'' DM values, as obtained by fitting independent ``DMX'' parameters at each epoch, are plotted in red. The dashed vertical lines delimit the one-year dip window. As apparent in the plot insets, the GP+dip model captures the transient feature more accurately. Furthermore, the GP-only model pays the flexibility required to fit the dip with higher variance across the entire data set. Both circumstances lead to higher predictive density for the GP+dip model; in this particular run, the delta log density is 39.}
\label{fig4}
\end{figure}

We perform two-fold cross validation 32 times with different random data partitions. The GP+dip model yields consistently higher predictive densities than the GP-only model, with delta log density \(54.8^{+584.2}_{-19.9}\) (quoted as median augmented by the 90\% interquantile range). The stronger predictive performance of the GP+dip model offers statistical evidence that the dip is a real physical feature.
This is confirmed by inspecting the reconstructed DM.
The gray bands in \autoref{fig4} show the 90\% interquantile range of reconstructed DM over 100 posterior draws of the model parameters in a single cross-validation run, while the red line traces the ``true'' DM at each epoch, as obtained by fitting independent dispersion values to multifrequency data.\footnote{In PTA jargon, the red line shows the best-fit ``DMX'' parameters.} It is apparent that the GP+dip model captures the DM transient more accurately, and that it follows the overall evolution of DM with smaller variance. Both conditions result in higher predictive density.

\section{Cross validation for GW detection in multi-pulsar datasets.}
To perform cross validation on multi-pulsar datasets and models (the latter possibly including a common GP describing the correlated delays induced by GWs for each pulsar), we may proceed without change \emph{if} we satisfy the representation condition: that is, if we partition the multi-pulsar residual vector, $Y$, into testing datasets $Y^{(k)}$ in such a way that all parameters of every pulsar are represented in every training dataset $Y^{(-k)}$. However, it seems natural to partition $Y$ instead into subsets that correspond to individual pulsars, or groups of individual pulsars. Such an arrangement may allow us to identify pulsars contaminated by pathological observations, pulsars that are poorly described by the noise model chosen for them, or pulsars that are too noisy to contribute to GW inference. 
We next discuss how to proceed for this more general partitioning. We describe separately the case of deterministic and stochastic GWs.

For GWs described by \emph{deterministic} models (e.g., isolated supermassive black-hole binaries), individual pulsars are described by the likelihoods of Eqs.\ \eqref{eq:hierarchical} and \eqref{eq:marginalized}, with the replacement $y \rightarrow Y^{(a)} - d^{(a)}(\theta_\mathrm{GW})$, where $Y^{(a)}$ is the vector of timing residuals for pulsar $a$, where the $\theta_\mathrm{GW}$ describe the GW parameters (a common set for all pulsars), and where the $d^{(a)}(\theta_\mathrm{GW})$ are the delays induced by the GWs on pulsar $a$. 
In this case, cross validation would proceed as follows: a) for each $Y^{(-k)}$ we would sample the joint posterior distribution of the $\theta_\mathrm{GW}$ and of the noise hyperparameters $\eta_{N,\mathrm{GP}}^{(-k)}$ describing the pulsars represented in $Y^{(-k)}$; b) for each corresponding $Y^{(k)}$, we would evaluate the log \emph{marginalized} predictive likelihood
\begin{multline}
\label{eq:cvdeterministic}
\log \int p(Y^{(k)}|\eta^{(k)},\theta_\mathrm{GW}) p(\theta_\mathrm{GW}|Y^{(k)}) p(\eta^{(k)}) \,
\mathrm{d}\eta^{(k)} \,
\mathrm{d}\theta_\mathrm{GW} \\
\simeq \log \frac{1}{N} \sum_{i=1}^N \int p(Y^{(k)}|\eta^{(k)},\theta^{(-k)}_\mathrm{GW,i}) p(\eta^{(k)}) \mathrm{d}\eta^{(k)},
\end{multline}
where the $\theta^{(-k)}_\mathrm{GW,i}$ are Markov Chain (sub-)samples from the training posterior $p(\theta_\mathrm{GW},\eta_{N,\mathrm{GP}}^{(-k)}|Y^{(-k)})$, and where we have dropped the $\eta_{N,\mathrm{GP}}$ suffix for compactness. From an implementation standpoint, the nested sum/integral in Eq.\ \eqref{eq:cvdeterministic} may require a dedicated stochastic algorithm similar to those employed to evaluate the Bayesian evidence \citep{2008ConPh..49...71T}.

It is important to notice that Eq.\ \eqref{eq:cvdeterministic} depends directly on the noise-hyperparameter priors $p(\eta^{(k)})$, which invalidates some of our motivation for computing predictive likelihoods in the first place. In practice, we may worry that we cannot compare predictive likelihoods for different $Y^{(k)}$ because they have different prior ``calibrations.''
With respect to this objection, it seems then natural to consider the ratio of Eq.\ \eqref{eq:cvdeterministic} to the noise-only evidence $\int p(Y^{(k)}|\eta^{(k)}) p(\eta^{(k)}) \, \mathrm{d}\eta^{(k)}$ (which in fact factorizes over the pulsars in $Y^{(k)}$).
We leave to future work the exploration of marginalized predictive likelihoods as GW detection statistics, as well as the development of an efficient sampling method for Eq.\ \eqref{eq:cvdeterministic}.

Last, for \emph{stochastic GWs} described by their spectrum and by their correlations across pulsars, we need to account not only for the common GW parameters $\theta_\mathrm{GW}$, but also for the correlations between the GW GP weights in each pulsar. To sketch the mathematical structure of the problem with more readable notation, let us consider the case of training on pulsar 1 and testing on pulsar 2; formulas generalize readily to $k$-fold--validation testing pairs $(Y^{(-k)},Y^{(k)})$.
We do as follows: We first obtain samples $(\eta^{(1)}_i,\theta_{\mathrm{GW},i})$ from the posterior
$p(\eta^{(1)},\theta_\mathrm{GW}|Y^{(1)})$;
we then evaluate the log marginalized predictive likelihood in the form
\begin{multline}
\label{eq:mplstochastic}
\log \sum_{i=1}^N \int p(Y^{(2)}|\eta^{(2)},c^{(2)}_\mathrm{GW})
p(c^{(2)}_\mathrm{GW}|c^{(1)}_\mathrm{GW},\theta_{\mathrm{GW},i})
\times \\
p(c^{(1)}_\mathrm{GW}|Y^{(1)};\eta^{(1)}_{i},\theta_{\mathrm{GW},i})
\, \mathrm{d}\eta^{(2)} \, \mathrm{d} c^{(2)}_\mathrm{GW}
\, \mathrm{d} c^{(1)}_\mathrm{GW}.
\end{multline}

The integral over the $c^{(1)}_\mathrm{GW}$ can be performed analytically, and the resulting conditional prior for the $c^{(1)}_\mathrm{GW}$ expressed as
\begin{equation}
p(c^{(2)}_\mathrm{GW}|Y^{(1)};\eta^{(1)}_{i},\theta_{\mathrm{GW},i})  =
\mathcal{N}(\bar{c}^{(2)|(1)},\Sigma^{(2)|(1)})
\end{equation}
with
\begin{equation}
\bar{c}^{(2)|(1)}(Y^{(1)};\eta^{(1)}_{i},\theta_{\mathrm{GW},i}) = \Phi_{21} \Phi^{-1}_{11} \bar{c}^{(1)},
\end{equation}
and
\begin{multline}
\Sigma^{(2)|(1)}(\eta^{(1)}_{i},\theta_{\mathrm{GW},i}) \\ =
\Phi_{22} - \Phi_{21}(\Phi_{11}^{-1} - \Phi_{11}^{-1} \Sigma^{(1)} \Phi_{11}^{-1}) \Phi_{21},
\end{multline}
where $\bar{c}^{(1)}(Y^{(1)};\eta^{(1)}_{i},\theta_{\mathrm{GW},i})$ and $\Sigma^{(1)}(\eta^{(1)}_{i},\theta_{\mathrm{GW},i})$ are given in Eqs.\ \eqref{eq:conditionalmean} and \eqref{eq:conditionalcov}, and the $\Phi_{ij}(\theta_{\mathrm{GW},i})$ denote the blocks of the joint normal prior for the GW GP weights.
The integral over the $c^{(2)}_\mathrm{GW}$ in Eq.\ \eqref{eq:mplstochastic} may also be performed analytically by way of Eq.\ \eqref{eq:marginalized}, with the replacement $y \rightarrow Y^{(2)} - F^{(2)}_\mathrm{GW}\bar{c}^{(2)|(1)}$ and $\Phi_\mathrm{GW} \rightarrow \Sigma^{(2)|(1)}$ (in this notation, $F^{(2)}_\mathrm{GW}$ and $\Phi_\mathrm{GW}$ represent the GW blocks of $F_\mathrm{GP}$ and $\Phi_\mathrm{GP}$).
Again, we leave to future work the investigation of marginalized predictive likelihoods as detection statistics for stochastic GWs in PTA data.

\section{Discussion}

The development of sophisticated noise models is of paramount importance to ongoing PTA searches for nanohertz GWs, especially so because much of the PTA ``instrument'' was not engineered by humans. Contrast LIGO's precisely fabricated arms with a PTA Earth--pulsar ``arm'', which comprises a radiotelescope, a distant ($\sim$ kpc) pulsar, and the expanse of spacetime in between. The telescope admits some experimental control, since we can test its signal response and mitigate noise in the receiver, but we still have to contend with poorly constrained physical processes in pulsar interiors and emission regions \citep{2015MNRAS.449.3293L}, not to mention dispersive effects as radio pulses propagate through the ionized interstellar medium to the Earth \citep{2010arXiv1010.3785C,2019arXiv190300426L,2018ApJ...861..132L,2016ApJ...821...66L}. Thus PTA noise models must be well motivated physically, and yet flexible enough to accommodate unknown unknowns, if they are to allow the identification of subtle GW-induced delays.

Current techniques to test the relative aptness and robustness of noise models include basic checks (i.e., evaluating $\chi^2$ fit residuals under different models), frequentist approaches (computing receiver operating characteristic curves to maximize the signal detection probability at a given false-alarm probability), as well as Bayesian model comparison \citep{2013PhRvD..87d4035T,2017PhRvD..95d2002T,2015PhRvD..91h4055S,2014PhRvD..90j4028T,2016PhRvD..93j4047C}.
In this last technique, we marginalize likelihoods for different models over their respective prior volumes, and use the ratios of marginal likelihoods (usually under the assumption of equal prior probabilities for each model) to make statements about the posterior odds with which one model is favored over the other. This approach is perfectly valid, but it can be troubled by practical issues such as accurately integrating into the tails of the likelihood distribution, as well as assigning the ranges of parameter priors in the first place. 

Bayesian cross validation addresses some of these issues by partitioning datasets into training and testing samples. Models are conditioned on the training set, producing posterior probability distributions for the parameters, over which we average the likelihood of the testing set. The cross-validation score is then the probability of the test data under the trained model. For reasonably informative training data, posteriors will be more compact than the priors, shielding the integration from ad hoc prior choices. To evaluate the integrals, it is convenient to average the likelihood over posterior samples from a conventional MCMC analysis of the training set.

In this article we argued for the power of Bayesian cross validation as applied to PTA data. Our case studies included the spectral characterization of a GW background, and the modeling of dispersive noise from the interstellar medium. For the former, we performed two-fold cross-validation on a dataset from the $1^\mathrm{st}$ IPTA mock data challenge \citep{2018arXiv181010527H}, showing that the data favored a correlated process consistent with a stochastic GW background from supermassive binary black holes. For the latter, two-fold cross-validation on $11$ years of NANOGrav data for pulsar J$1713$+$0747$ illustrated the necessity to include a transient dispersive noise feature around the year $2009$, consistent with a void in the electron density along the line of sight \citep{2016MNRAS.458.2161L}. We also introduced a formalism for multi-pulsar cross validation, where GW models conditioned on training data from a subarray are assessed for predictive performance on left-out pulsars. In future work we will investigate this approach as a tool to validate claims of GW detections in real PTA datasets.

We expect cross validation to be similarly useful in analyzing data from other GW detectors. For instance, within the global network of ground-based GW interferometers, support for a GW signal in one detector could be validated using data from a different widely separated detector. Furthermore, signals such as GW150914, which could have been observed by a LISA-like detector years before it was seen by LIGO \citep{2016PhRvL.116w1102S}, raise the prospect of multi-band, multi-detector cross-validation of GW signals.

\vspace{12pt}

\paragraph*{Acknowledgments.}
We thank Joseph Simon and Michael Lam for discussions regarding the dispersion-measure variation of PSR J$1713$+$0747$. This research
was performed in part using the Zwicky computer cluster at
Caltech supported by NSF under MRI-R2 award No. PHY0960291 and by the Sherman Fairchild Foundation. Portions of this research were carried out at the
Jet Propulsion Laboratory, California Institute of Technology,
under a contract with the National Aeronautics and Space
Administration. This work
was supported in part by National Science Foundation Grant
No. PHYS-1066293 and by the hospitality of the Aspen Center for Physics. MV was supported by the Jet Propulsion Laboratory RTD program.
SRT was supported by the NANOGrav National Science Foundation Physics Frontier Center, award number 1430284. Parts of this work were carried out at the Jet Propulsion Laboratory, California Institute of Technology, under contract to the National Aeronautics and Space Administration. Copyright 2019 California Institute of Technology. Government sponsorship acknowledged.

\bibliographystyle{mnras}
\bibliography{bib}

\end{document}